\pdfoutput=1
\documentclass[a4paper,12pt]{article}
\def\Vbg{V_{{G}}}

\def\={\,=\,}

\def\Vbg{V_{G}}
\def\rc{{r_{C}}}
\def\idc{{I_{DC}}}
\def\kf{\mathrm{k_{F}}}

\usepackage{amssymb}
\usepackage{amsmath}
\usepackage{graphicx}
\usepackage{graphics}
\usepackage{color}
\usepackage[square,sort,comma,numbers,super,sort&compress]{natbib}
\usepackage{multirow}
\usepackage{upgreek}
\usepackage[applemac]{inputenc}
\usepackage{upgreek}
\usepackage[normalem]{ulem}
\usepackage{newfloat}
\usepackage{float}
\usepackage{setspace}
\usepackage{soul}
\usepackage{caption}
\usepackage{titling}
\usepackage{authblk}
\usepackage{geometry}
\DeclareFloatingEnvironment[name={Supplementary Figure}]{suppfigure}

\title{Supercurrent in the quantum Hall regime}

\author[1,2*]{F. Amet} 
\author[1*]{C. T. Ke} 
\author[3]{I. V. Borzenets}
\author[1]{Y-M. Wang}
\author[4]{K. Watanabe}
\author[4]{T. Taniguchi}
\author[5]{R.S. Deacon}
\author[3,6]{M. Yamamoto} 
\author[1]{Y. Bomze} 
\author[3,5]{S. Tarucha}
\author[1]{G. Finkelstein}

\affil[1]{Department of Physics, Duke University, Durham, NC 27708, USA.} 
\affil[2]{Department of Physics and Astronomy, Appalachian State University, Boone, NC 28607, USA.}
\affil[3]{Department of Applied Physics, University of Tokyo, Bunkyo-ku,Tokyo, 113-8656, Japan.}
\affil[4]{Advanced Materials Laboratory, National Institute for Materials Science, 1-1 Namiki, Tsukuba, 305-0044, Japan.}
\affil[5]{Center for Emergent Matter Science (CEMS), RIKEN, Wako-shi, Saitama 351-0198, Japan.}
\affil[6]{PRESTO, Japan Science and Technology Agency, Kawaguchi-shi, Saitama 332-0012, Japan.}

\begin{document}
	\date{}	
	\maketitle	

	\begin{abstract}
		
A novel promising route for creating topological states and excitations is to combine superconductivity and the quantum Hall (QH) effect~\cite{Lindner_2012,Clarke_2013}. Despite this potential, signatures of superconductivity in the quantum Hall regime remain scarce~\cite{Takayanagi_98,Moore_99,Rokhinson2015, Gueron2011,Schonenberger2012,XuDu,Vandersypen2015}, and a superconducting current through a QH weak link has so far eluded experimental observation. Here we demonstrate the existence of a new type of supercurrent-carrying states in a QH region at magnetic fields as high as  2$\,$Tesla. The observation of supercurrent in the quantum Hall regime marks an important step in the quest for exotic topological excitations such as Majorana fermions and parafermions, which may find applications in fault-tolerant quantum computations.		
		
\end{abstract}
\newpage

The interplay of the quantum Hall effect with superconductivity is expected to result in novel excitations with non-trivial braiding statistics such as Majorana fermions and non-abelian Majorana anyons~\cite{Lindner_2012,Clarke_2013,Mong2014,San-Jose2015}. 
When a quantum Hall region is contacted by two superconducting electrodes, the gapped QH bulk prevents the flow of a supercurrent. 
However, it was predicted more than 20 years ago that the supercurrent may still be mediated by QH edge states~\cite{Zyuzin1993}. 
Due to its chiral nature, a single edge can only conduct charge carriers in one direction, so both edges have to be involved in establishing supercurrent between the two contacts. This situation is fundamentally different from the Josephson junctions made of two-dimensional topological insulators, where each edge can support its own supercurrent\cite{Knez2012,Yacoby2014,Kouwenhoven2015,Molenkamp2015}. 
Indeed, contrary to the case of topological insulators, the magnetic field in the QH regime breaks time-reversal symmetry, which is essential for the s-wave pairing of conventional superconductors. Nonetheless, we observe a robust supercurrent in the quantum Hall regime, which we attribute to an unconventional form of Andreev bound states circulating along the perimeter of the QH region and involving electron and hole trajectories separated by several micrometers. 

We performed transport measurements on four Josephson junctions (J$_{1-4}$) made of graphene encapsulated in boron nitride and contacted by electrodes made of a molybdenum-rhenium alloy [Fig. 1a]~\cite{Vandersypen2015}, a type II superconductor with a high upper critical field of H$_{c2}$=8 T. The high quality of these heterostructures allowed us to observe Fabry-Perot oscillations of the junctions' resistance and critical current, indicating that the transmission of charge carriers between the contacts is ballistic~\cite{Suppinfo}. The supercurrent is uniformly distributed along the width of the contacts, as evidenced by the regular Fraunhofer pattern~\cite{Tinkham} measured at small magnetic fields~\cite{Suppinfo}. All junctions demonstrate supercurrent in the QH regime; for consistency, we choose to present data measured on sample J$_1$, which has a distance between contacts $L$$\,=\,$0.3$\,\mu$m and a width of the contacts $W$$\,=\,$2.4$\mu$m (see Figure 1b). 

Recent preprint reported on the observation of supercurrent through encapsulated graphene in moderate magnetic fields, when the diameter of the cyclotron orbit is larger but comparable to the length of the junction, $2\rc \geq L$~\cite{Geim2015}. (Here, $\rc$$\,=\,$$\hbar k_F/eB$ is the cyclotron radius.) This supercurrent has been attributed to Andreev bound states made of closed trajectories connected by several elastic and Andreev reflections, which yield pockets of superconductivity at random values of density and field. We further explore this regime in the Supplementary Information. In the main text, we demonstrate that a completely new regime emerges at even larger magnetic fields, when $\rc$ is much smaller than the device dimensions and the mean free path. In this regime, the bulk of the junction is gapped by Landau quantization so that a current may only flow along the edges. 

\begin{figure*}[b!]
\center \label{fig2}
\includegraphics[width=170mm]{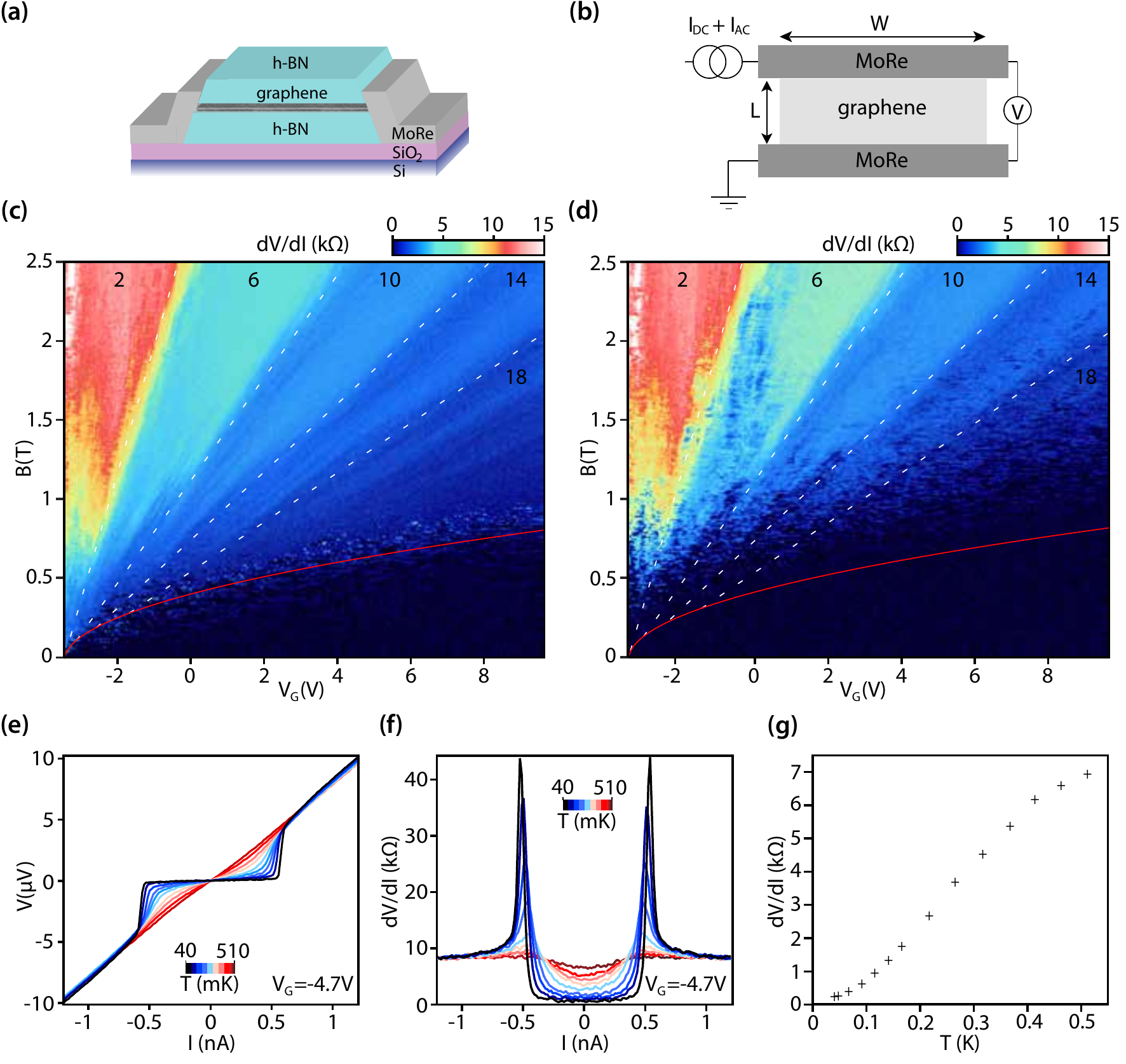}
\caption{(a) Schematic of the encapsulated graphene heterostructure with molybdenum rhenium contacts. (b) Schematic of the measurement setup. (c-d) Fan diagrams of the differential resistance $dV/dI$ plotted vs. gate voltage $\Vbg$ and magnetic field $B$. Panel (c) is measured at a finite current bias of $I_{DC}=6$ nA which suppresses superconductivity in the QH regime and reveals the quantized plateaus. Panel (d) is measured at zero DC current and shows superconducting pockets extending beyond the semiclassical parabolic region of $2\rc \geq L$. (e)  $I-V$ curves measured in a superconducting pockets at $B=1$ T and the filling factor in the range of $\nu \approx 2$. The supercurrent branch is clearly visible at the lowest temperature (40 mK) for $I$$\,<\,$0.5 nA. (f) The temperature dependence of the corresponding differential resistance, $dV/dI$. The resistance reaches maximum at $I_S$$\,=\,$0.5 nA, at which point the junction switches from the superconducting to the normal branch. (g)  $dV/dI$ measured as a function of temperature at $I=0$, showing gradual suppression of superconductivity at elevated temperatures due to the phase diffusion.}
\end{figure*}

Figures 1c and d show the differential resistance of the sample, $R\equiv dV/dI$, plotted vs. back gate voltage, $\Vbg$, and magnetic field, $B$. The resistance is measured in a four-terminal configuration where four MoRe electrodes merge into two contacts on each side of the junction (Figure 1b). The map in Figure 1c is measured with an AC excitation current $I_{AC}$$\,=\,$50$\,$pA applied on top of a large DC current of $I_{DC}=6\,$nA, which suppresses supercurrent and highlights the QH features. As $B$ increases, a fan diagram characteristic of the quantum Hall effect in graphene emerges: resistance plateaus follow contours of constant filling factor $\nu\equiv\frac{nh}{eB}$$\,=\pm \,$2, 6, 10,$\ldots$~\cite{Katsnelson} This quantization becomes visible as soon as $B$ exceeds the red parabolic contour $2\rc=L$ because device dimensions prevent the development of the quantum Hall effect at lower fields. The dark region under the parabola (2$\rc >L$) indicates a vanishing differential resistance as a supercurrent of tens of nA may flow in this semiclassical regime~\cite{Geim2015}.

Figure 1d shows $R(\Vbg,B)$ measured simultaneously with Figure 1c using exactly the same AC excitation of 50 pA, but without applying a DC current. Strikingly, pockets of supercurrent extend far into the quantum Hall regime. They are visible as dark spots of vanishing resistance above the parabolic contour. These pockets occur at somewhat random values of $\Vbg$, but are highly reproducible as the gate voltage is swept back and forth. 
To check that these regions do indeed correspond to a supercurrent, in Figure 1e we show the $I - V$ curves measured in one of the superconducting pockets at at $B$$\,=\,$1$\,$T and $\Vbg = - 4.7$V.  The curves demonstrate a clear supercurrent branch, which extends up to $I$$\,<\,$0.5 nA at the lowest temperature of 40 mK. We stress that at that particular point, $\rc \approx 25$$\,$nm $\ll L/2=150$ nm. The corresponding differential resistance ($dV/dI$) vanishes in the same range of currents (Figure 1f). The maximum of resistance is reached at $I_S$$\,=\,$0.5 nA, at which point the sample switches from the superconducting to the normal branch. 

The curves in Figures 1e,f are extremely sensitive to temperature, the supercurrent being washed away by $T\sim 500\,$mK. This energy scale is orders of magnitude below the critical temperature of MoRe ($\approx\,$10$\,$K) and the energy splitting of the lowest Landau levels in graphene ($>$100 K). It is however close to the Josephson energy, $E_{J}=\hbar I_{C}/2e$, which is tens of mK for critical currents of a few nA. For temperatures comparable to the Josephson energy, the apparent switching current, $I_S$, is expected to be suppressed by thermal fluctuations with respect to the true critical current $I_{C}$. This explains the observed $I_S$ of only 0.5$\,$nA in Figure 1f. The thermal fluctuations also result in phase diffusion~\cite{Tinkham}, which yields a finite junction resistance even at zero DC current (Figure 1g).

\begin{figure*}[b!]
\center \label{fig3}
\includegraphics[width=90mm]{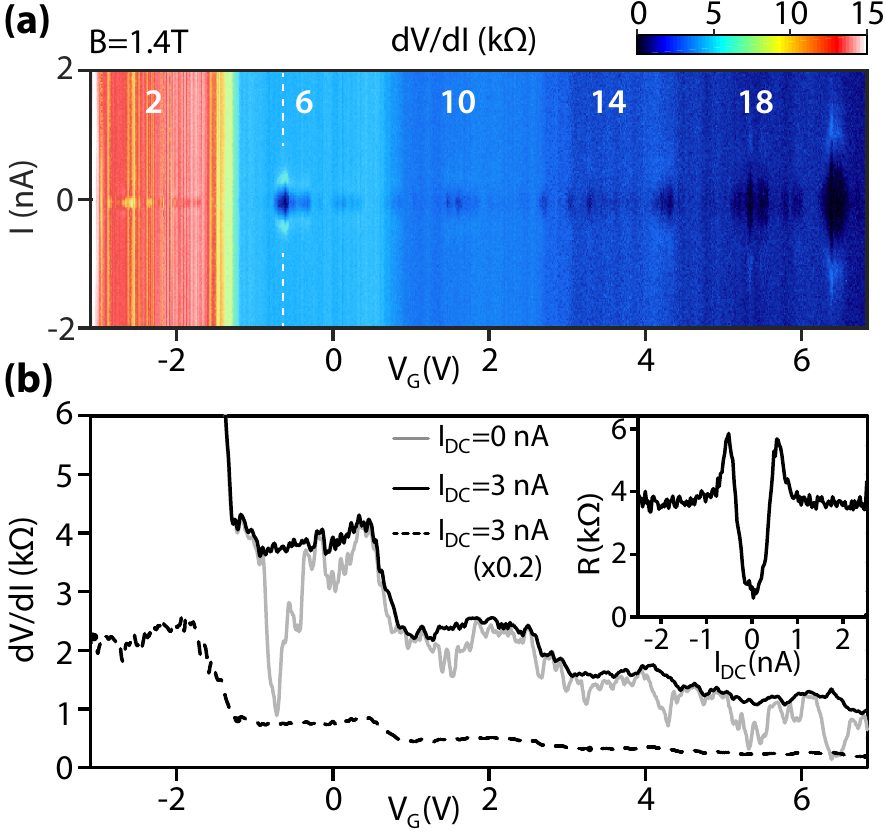}
\caption{(a) Differential resistance $dV/dI$ measured at 45$\,$mK as a function of bias current $I$ and gate voltage $\Vbg$ at constant $B=1.4\,$T. Filling factors are indicated in bold white font on panel (a). QH plateaus are clearly visible as stripes of different colors around filling factors $\nu=4(n+1/2)$ with an integer $n$. Pockets of superconductivity appear as dark regions close to zero current. (b) Line cuts of $dV/dI$ in the same range of $\Vbg$, measured at zero DC current (gray) and at $I_{DC}=3$ nA (black) applied to suppress the supercurrent. The dashed curve is identical to the black one, but scaled by a factor 0.2 to demonstrate the $\nu$$\,=\,$2 plateau. Inset: $dV/dI$ vs. $I$ for one of the prominent superconducting pocket at $\nu=6$ indicated in panel (a) by a vertical dashed line. }
\end{figure*}

To further illustrate the coexistence of the QH and the superconducting pockets, we show the differential resistance of the same junction measured as a function of $\Vbg$ and the current bias $I$ at $B$$\,=\,$1.4$\,$T (Figure 2a). The QH plateaus are visible in Figure 2a as vertical stripes of different color. Pockets of superconductivity appear around zero bias as dark minima of $dV/dI$. (At this field, the cyclotron radius $\rc$$\,\approx\,$15$\sqrt{\nu}$$\,$nm is much smaller than device dimensions throughout the map.) The solid black line in Figure 2b shows the cross-section of the $dV/dI$ map taken with a finite current bias of $I_{DC}\,=\,3$ nA, high enough to suppress any superconducting features. Plateaus are clearly visible close to quantized values of $R$$\,=\,$$h/(\nu e^2)$, with $\nu$$\,=\,$2, 6, 10, $\ldots$ The deviations from perfect quantization are common in two-probe measurements~\cite{Jimmy}. The gray line corresponds to the cross-section measured at zero DC current, which clearly shows the regions of suppressed differential resistance formed on top of the plateaus due to superconductivity.

Figures 3a-c show the differential resistance measured at three superconducting pockets as a function of the bias current and magnetic field, which is varied in steps of $0.1$ mT around $B=1$ T. The critical current exhibits a robust interference pattern as the magnetic field is varied, with a period of $0.5$$\,$mT. Remarkably, this value is close to the period of the Fraunhofer pattern measured on the same junction at very low fields ($B<10$mT), when the current distribution along the width of the junction is uniform (see Figure S3). However, the current distribution becomes spatially inhomogeneous in the intermediate magnetic fields of tens of mT and beyond, resulting in a very irregular pattern of the supercurrent vs. magnetic field (see Ref.~\cite{Geim2015} and Figure S9). Therefore, the periodicity recovered at high field must be attributed to a very different mechanism.  

Since at 1$\,$T the bulk of graphene is clearly gapped, the periodic supercurrent must be mediated by the edge states. However, the edge states with opposite momenta are located on the opposite sides of the sample, separated by 2.5 to 4.5 $\mu$m in our junctions. This scale greatly exceeds the coherence length of the MoRe electrodes (a few nanometers), which prevents the direct coupling of the edges through a simple Andreev reflection. Below, we discuss a mechanism that couples the edge states through the hybrid electron-hole modes which are formed at the interfaces between the superconducting contacts and the QH region~\cite{Takagaki1998,Hoppe}. 

Due to the pairing gap, single particles cannot enter the superconducting electrodes and have to form edge states along the superconductor-QH interfaces (Figure 3d). The electron and hole states propagate in the same direction and are hybridized by the superconducting proximity, resulting in chiral hybrid modes. Quasiclassically, one can picture the hybrid electron-hole mode as a skipping orbit in which an electron and a hole are converted into one another on each bounce from the superconductor~\cite{Hoppe} (Figure 3d). Depending on the transparency of the interface, such a mode could have various degrees of mixing between its electron and hole components. In particular, for perfectly transparent interfaces, it becomes a neutral mixture of the two carrier types, similar to the Majorana modes. 

\begin{figure*}[t!]
\center \label{fig4}
\includegraphics[width=170mm]{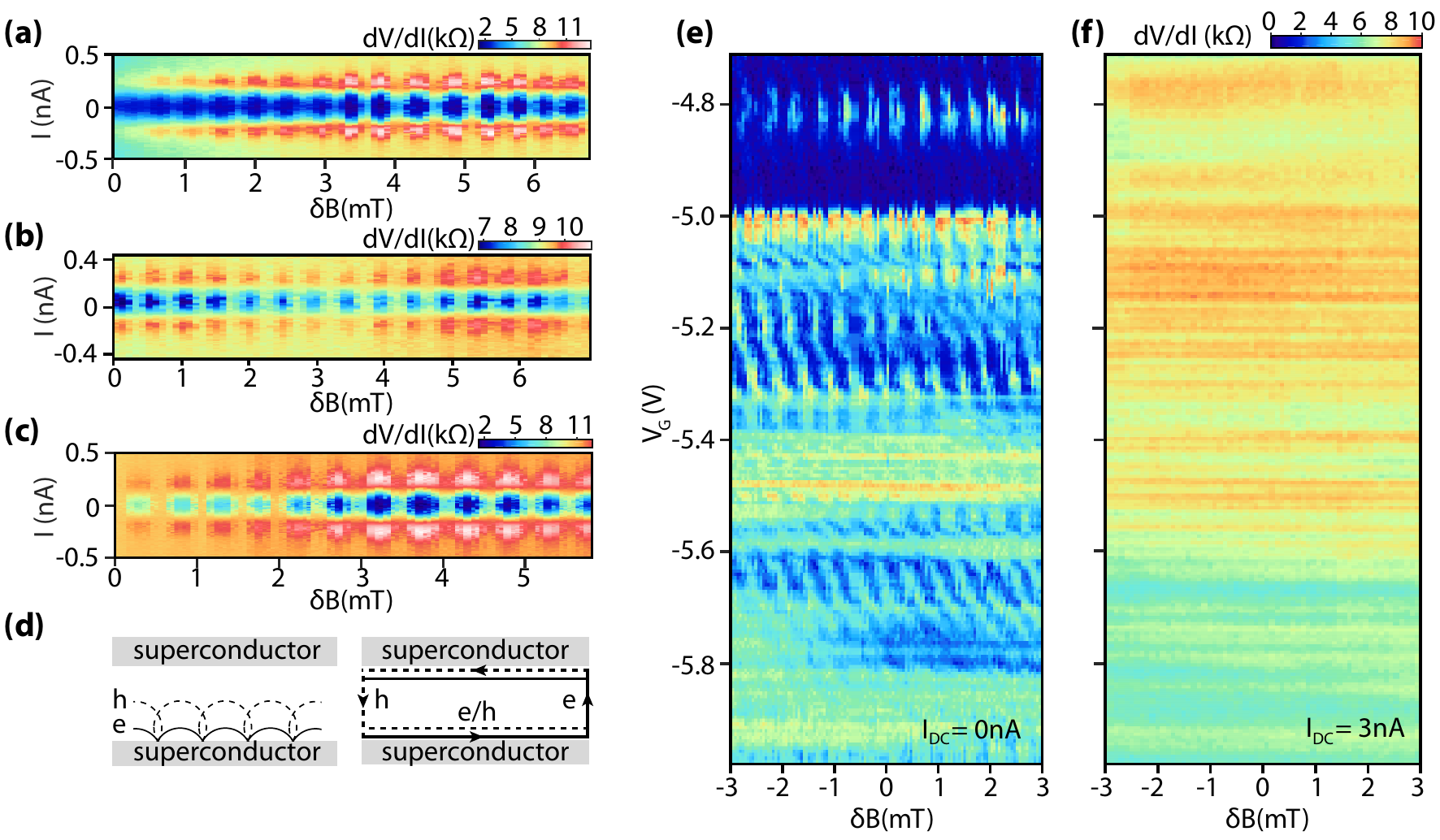}
\caption{(a-c) Differential resistance $dV/dI$ measured at 40 mK as a function of bias current $I$ and incremental magnetic field $\delta B$ around $B=1\,$T. The three panels correspond to the superconducting pockets which are located at gate voltages of (a) $-5.1\,$V (b) $-2.2\,$V and (c) $-2.6\,$V. In all three cases, the critical current oscillates with the same period of 0.5 mT, which is close to the period of the Fraunhofer pattern at small fields (Figure S3). (d) Right: schematics of the Andreev bound states, made of the electron and hole states on the opposite sides of the sample, which couple through the hybrid electron-hole modes running along the superconductor-QH interfaces. Left: these chiral hybrid modes are made of electron and hole edge states mixed through the Andreev processes, as described in the main text. (e,f) Maps of $dV/dI$ as a function of $\delta B$ and $\Vbg$, measured at (e) zero $I_{DC}$ and (f) $I_{DC}=$3 nA when the supercurrent is suppressed. Clearly, the superconducting features of panel (e) demonstrate the same magnetic field periodicity of $0.5$ mT as found in panels (a-c), while the normal resistance in panel (f) is almost field-insensitive in this field range.}
\end{figure*}

The hybrid modes provide a coherent reservoir of correlated electrons and holes, which is spread over the length of the superconducting electrode and thus could couple the edge states on the opposite sides of the sample~\cite{Beenakker2011} (Figure 3d). Specifically, an electron approaching the top contact along the right edge of graphene must be converted to the hybrid electron-hole mode, which then propagates along the graphene-superconductor interface to the left. Here, it has a finite amplitude of coupling to the left QH edge state as a hole, which then flows to the bottom contact. The loop is completed by the hole conversion into the hybrid mode at that contact, and by its subsequent coupling to the original electronic state at the bottom right corner~\cite{Beenakker2011}. This mechanism shuttles one Cooper pair between the contacts, coupling in the process the single-particle edge states separated by microns. Note that the process of an electron entering and a hole leaving the hybrid mode may be viewed as a perfect crossed Andreev reflection over distances of several microns.

To substantiate this scenario, we study the dependence of the superconducting features on the magnetic field and the gate voltage (Figure 3e,f). Here, panel (f) is measured at $I_{DC}=3$ nA, exceeding the supercurrent, while panel (e) is measured at zero DC current and shows suppressed resistance when supercurrent flows between the contacts. Clearly, the normal features in Figure 3f are almost field-insensitive, while the superconducting features in Figure 3e exhibit the same magnetic field periodicity as in Figures 3a-c. Remarkably, the phase of these features depends on $V_{G}$. 

Indeed, the quantized phase of Andreev bound states in panel (e) is made of two contributions: the Aharonov-Bohm phase due to the magnetic flux through the junction, and the phase accumulated by carriers completing the loop trajectory. The first term yields the magnetic field periodicity of Figure 3a-c and 3e. The second term is determined by the carrier momentum and therefore depends on $V_{G}$. To keep the total phase constant, the contributions of the two phases have to cancel, resulting in the diagonal contours of constant phase in Figure 3e. 
Note that Figure 3e rejects a hypothetical alternative scenario in which each edge would support a separate superconducting path, like in the case of 2D topological insulators. Indeed, their interference phase would only depend on $B$ and not on $\Vbg$, resulting in the decoupling of the gate voltage and magnetic field dependencies. This would give rise to vertical strips in Figure 3e, contrary to our observations.

The nature of hybrid mode propagating along the superconducting interface likely explains the extreme sensitivity of the supercurrent to $\Vbg$. Indeed, mesoscopic details of the superconducting interface (such as the contact roughness, fluctuations in the interface transparency, or the presence of disorder) should strongly affect the relative phase and amplitudes of the electron and hole components of the hybrid mode. These in turn determine the coupling of the hybrid mode to the QH edge states at the corners of the sample (Figure 3d), likely resulting in the mesoscopic fluctuations of the superconducting current as $\Vbg$ is varied.

In conclusion, we have measured supercurrent through a quantum Hall region, which is mediated by the Andreev bound states encompassing the edge channels on the opposite sides of the sample. These states are decidedly noninvariant under time reversal, and observing them makes an important step toward the realization of artificial superconducting hybrids in the quest for Majorana fermions and other exotic topological excitations proposed in Refs.~\cite{Lindner_2012,Clarke_2013,Mong2014}. Fractional quantum Hall states in graphene~\cite{FQHE1, FQHE2}, which emerge in fields as small as 5$\,$T~\cite{Amet}, should bring these proposals into the realm of possibility. We also anticipate that control of these excitations will be greatly facilitated in 2D graphene nanostructures, where edge channels can be easily manipulated, split, and combined by the application of gate voltages. 

\paragraph{Acknowledgments}

We thank Harold Baranger, Patrick Gallagher, Konstantin Matveev, Karen Michaeli, Alex Smirnov, James Williams, and Sergey Yarmolenko for fruitful discussions and technical help. The work in the US was supported by the Division of Materials Sciences and Engineering, Office of Basic Energy Sciences, U.S. Department of Energy, under Award No. DE-SC0002765. F.A. acknowledges support from the Fritz London postdoctoral fellowship. I.V.B. is funded by the Canon foundation and Grants-in-Aid for Scientific Research on Innovative Areas, Science of Atomic Layers. R.S.D. is funded by Grants-in-Aid for Young Scientists B ($\#$26790008).

\clearpage
\newpage
\title{Supplementary information for \\ "Supercurrent in the quantum Hall regime"}

\author[1*]{F. Amet} 
\author[1*]{C. T. Ke} 
\author[2]{I. V. Borzenets}
\author[1]{Y-M. Wang}
\author[3]{K. Watanabe}
\author[3]{T. Taniguchi}
\author[4]{R.S. Deacon}
\author[2,5]{M. Yamamoto} 
\author[1]{Y. Bomze} 
\author[2,4]{S. Tarucha}
\author[1]{G. Finkelstein}
\date{}	
\maketitle	

\newpage
	\section{Device Fabrication}
	
	Graphene and boron nitride flakes are exfoliated on separate silicon wafer pieces with a 300$\,$nm-thick thermally grown oxide without prior oxygen plasma treatment.  A 2$\times$2$\,$mm piece of polydimethylsiloxane (PDMS) is then adhered on a glass slide and treated in oxygen plasma for 1$\,$min (60W, 250mTorr). A 1$\,$$\mu$m thick film of polypropylene carbonate (PPC) is spin-coated on a bare piece of a silicon wafer and baked at 80$^{\circ}$C for 10 minutes. The PPC film is then mechanically peeled from silicon, which is facilitated by thicker PPC edge beads near the substrate edges, forming a relatively rigid frame. The PPC film is then carefully deposited on the PDMS stamp and baked 10$\,$min at 80$^{\circ}$C. 
	In order to pick up flakes from their original substrate, the PDMS stamp is brought into contact with the flake as slowly as possible and baked at 50$^{\circ}$C; the flake is then usually picked up by the stamp as it is lifted. In order to deposit the assembled stack on the final substrate, the stamp is then baked at 90 to 110$^{\circ}$C and peeled off very slowly as the stack stays on the substrate. 
	
	The resulting stacks usually have smooth defect-free terraces over several microns, separated by bubbles of trapped adsorbates$^{[\mathrm{S}1]}$. When the distribution of bubbles is not suitable for the fabrication of a defect-free device, we found that mild heating of the stacks at temperatures ranging from 200$^{\circ}$ to 250$^{\circ}$C often allows bubbles to migrate and rearrange into a different, more favorable distribution as terraces of BN/Graphene/BN ``self clean''. It is critical that the final mesa of the device is positioned in the middle of a terrace free of defects as evidenced by Raman spectroscopy [Fig. S1b].

'Electrical contacts to the flakes are patterned using e-beam lithography. The superconducting contacts are patterned with relatively thick PMMA (450$\,$nm), then reactive-ion etched in a CHF$_{3}$/O$_{2}$ mixture (flow rates 40/6 sccm) at 1$\,$Pa and 60$\,$W power. Etch time varies between 90 and 210 seconds depending on the initial thickness of the stack. 	

	\begin{suppfigure}[h!]
		\center \label{fig1}
		\includegraphics[width=172mm]{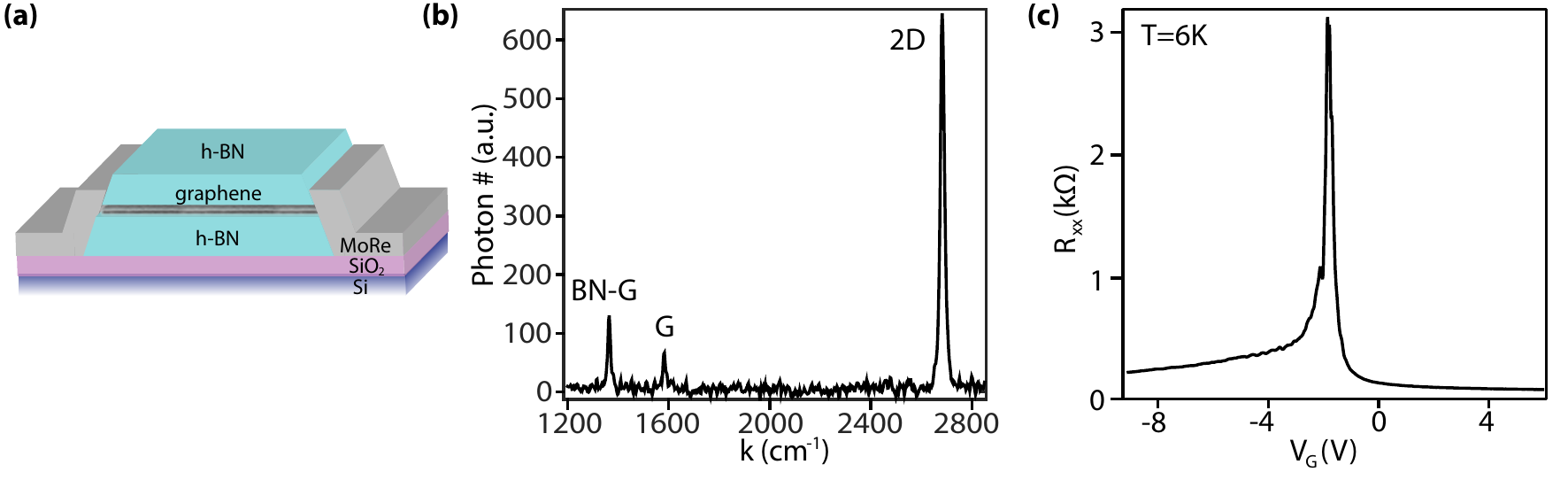}
		\begin{spacing}{1.0}
			\caption{(a) Schematic of the device: graphene is encapsulated between boron nitride layers and contacts are made of molybdenum-rhenium alloy. A doped silicon back gate controls the carrier density in the sheet. (b) Raman spectrum of an encapsulated graphene flake prior to patterning, which shows the h-BN G-peak at 1360$\,$cm$^{-1}$, the graphene G peak at 1580$\,$cm$^{-1}$ and the 2D band at 2680$\,$cm$^{-1}$. The particularly large 2D/G peak ratio hints at the quality of the heterostructure. (c) Two terminal resistance of J$_3$ measured at 6K, above the critical temperature of the junction.}
		\end{spacing}
	\end{suppfigure}
	
	Superconducting contacts are then directly deposited using the same PMMA mask in a DC magnetron sputterer, which results in self-aligned quasi-one-dimensional contacts$^{[\mathrm{S}2]}$. The target consists of Molybdenum Rhenium alloy (50/50 wt$\%$) with 99.9$\%$ purity. The chamber pressure during sputtering reaches 2$\,$mTorr, with a power of 160$\,$W and a rate of approximately 50$\,$nm/min. A schematic of the final device is shown on Figure S1a. Table 1 lists the devices included in this study with their geometric parameters. Here, $L$ is the distance between the superconducting contacts, and $W$ is their width.

	\section{Evidence for ballistic transport at zero field}
	
	Figure S1c shows $dV/dI$$\,$($\Vbg$) for junction J$_3$ measured at 6$\,$K, above the critical temperature of the device. A narrow Dirac peak is visible at $\Vbg$$\,\approx\,$-2$\,$V, separating the electron and hole-doped regimes. The electron-hole asymmetry is typical for two-terminal resistance measurements and comes from the work function mismatch between the metal contacts and graphene; in our case, it yields n-doping of graphene at the contacts. 
	
	\begin{suppfigure}[b!]
		\center \label{fig2}
		\includegraphics[width=172mm]{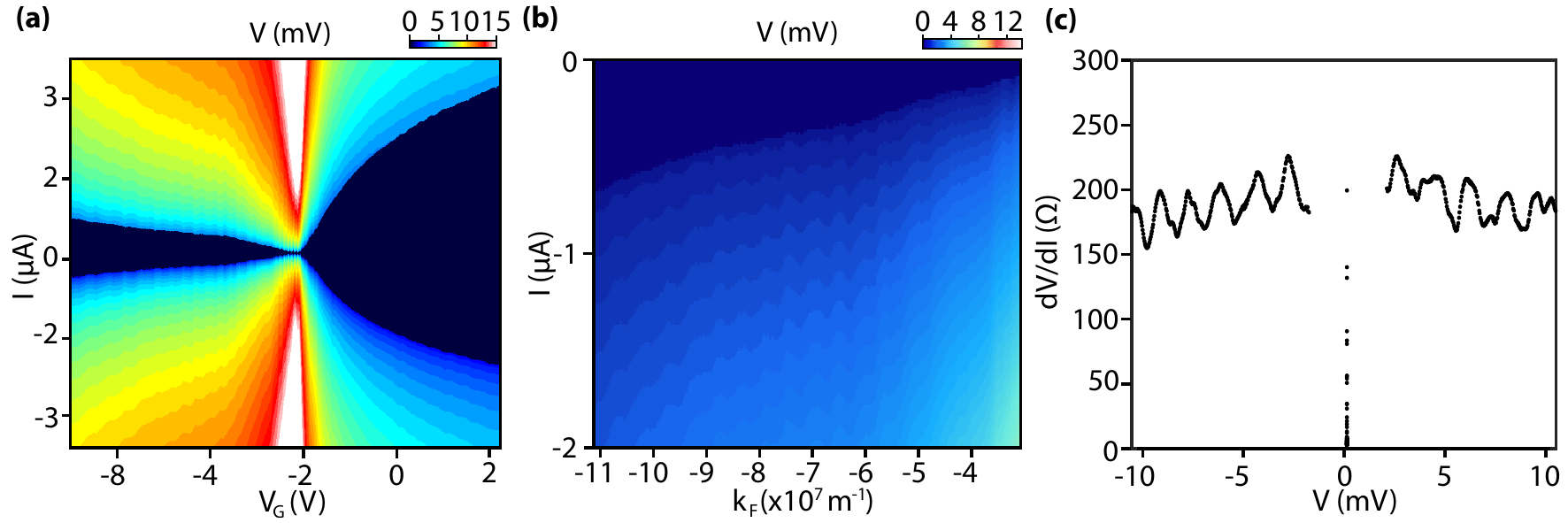}
		\begin{spacing}{1.0}
			\caption{(a) Voltage across the junction $\vert V(\Vbg,I)\vert$ measured at T$\,=\,$1.8K on J$_3$. (b) Close-up view of the Fabry Perot oscillations in the critical current from $\kf$$\,=\,$3$\times$10$^{7}$$\,$m$^{-1}$ to 11$\times$10$^{7}$$\,$m$^{-1}$. (c) Differential resistance $dV/dI$$\,$($V$) showing Fabry Perot oscillations as a function of bias $V$ in the normal state at $\Vbg$$\,=\,$-8.1V.}
		\end{spacing}
	\end{suppfigure}

	Figure S2a shows the voltage $V$ across the junction as a function of current bias $I$ and the gate voltage $\Vbg$ measured at $1.4$ K. Current flows without dissipation below the switching current $I_{S}$ (dark blue region indicating a vanishing voltage $V$ across the junction), beyond which point the junction switches to the normal state and a finite voltage appears. As expected, the switching current is minimal at the charge neutrality point. Furthermore, it is significantly smaller in the p-doped regime compared to the n-doped regime. The suppression is a result of the PN junctions formed close to the contacts. The partial reflections from the PN junctions also induces Fabry-Perot interference pattern, which can be observed in oscillations of the critical current  (Figure S2b). Oscillations are expected whenever the phase accumulated across the junction $k_{F}L$ is a multiple of $\pi$. We observe a periodicity $\Delta k_{F}\approx 6.4$$\times$$10^{6}\,$m$^{-1}$, in good agreement with the expected $4.8$$\times$$10^{6}\,$m$^{-1}$ for a 650$\,$nm-wide cavity. The measured $\delta k_{F}$ would correspond to an effective cavity length of 490$\,$nm. This length corresponds to the distance between PN junctions induced by the contacts in the hole doped regime and is therefore expected to be shorter than the actual length of the junction. The existence of these Fabry-Perot oscillations suggests that electrons contributing to supercurrent travel ballistically, as observed recently in short Josephson junctions$^{[\mathrm{S}2-\mathrm{S}5]}$. The resonant transmission of charge carriers through the cavity is also observed in the bias dependence of $dV/dI$, similar to Ref. S4 [Fig. S2c].

	\section{Homogeneity of the current distribution}
	
	\begin{suppfigure}[h!]
		\center \label{fig3}
		\includegraphics[width=172mm]{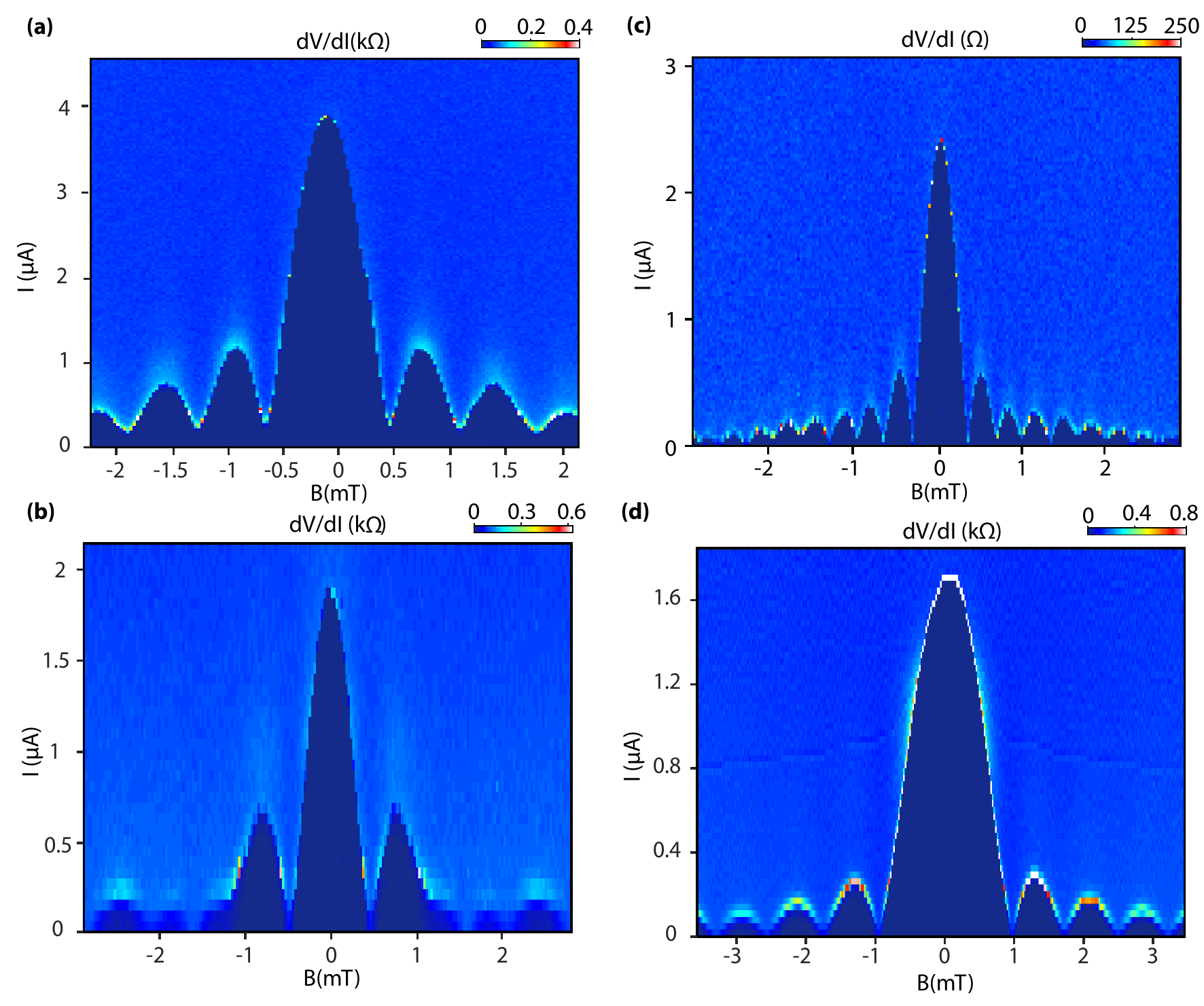}
		\begin{spacing}{1.0}
			\caption{a-d: Fraunhofer patterns for junctions J$_{1-4}$, measured at $T$$\,=\,$30$\,$mK in the electron-doped regime: $\Vbg$$\,=\,$6$\,$V (J$_{1-3}$) and $\Vbg$$\,=\,$3.1$\,$V (J$_{4}$). }
		\end{spacing}
	\end{suppfigure}
	
	As a small magnetic field is applied perpendicular to the plane of the Josephson junction, the conventional Fraunhofer-like interference pattern appears in $I_{C}(B)$, indicating that the supercurrent is uniformly distributed across the width of the junction (Figure S3)$^{[\mathrm{S}6]}$. The critical current vanishes whenever an integer multiple of magnetic flux quanta $\Phi_{0}\equiv h/2e$ are threaded through the junction. The flux though graphene is enhanced due to flux focusing by the superconducting leads$^{[\mathrm{S}7]}$, which in our case are wider than the graphene region. We can estimate the effective flux focusing area as the area of the MoRe region that is closer to the graphene interface than to any other edge. Depending on the width of the the two contacts $C_{1,2}$, this area is on the order of $W^2/2$ or $W\times (C_1/2+C_2/2)$, (whichever is smaller). Once flux focusing is included, the expected periodicity is close to our observations, summarized in Table 1.
	\begin{table*}[h!]
		\begin{center}
			\begin{spacing}{1.0}
				\caption{List of samples}
				\label{tab:table1}
				\begin{tabular}{|c|c|c|c|c|c|}
					\hline
					Device name & Length & Width & Estimated Flux & Expected & Measured\\ 
					& & &Focusing Area & period & period\\
					\hline
					&  &  &  &  &\\
					$J_{1}$  & 0.3$\,$$\mu m$ & 2.4$\mu m$ & 2.9$\mu m^2$ & 0.6 mT & 0.6 mT\\
					$ J_{2}$  & 0.8$\,$$\mu m$ & 2.4$\mu m$& 2.3$\mu m^2$ & 0.5 mT & 0.7 mT\\
					$J_{3}$  & 0.65$\,$$\mu m$ & 4.5$\mu m$ & 3.7$\mu m^2$ & 0.3 mT & 0.3 mT\\
					$J_{4}$ & 0.5$\,$$\mu m$ & 2.7$\mu m$& 2$\mu m^2$ & 0.6 mT & 0.8 mT\\   
					\hline
				\end{tabular}
			\end{spacing}
		\end{center}
	\end{table*}

\section{Fan diagrams and superconductivity}

	Figure S4 explicitly shows the location of our measurements performed on junction $J_{1}$ on top of the QH diagram. The diagram itself is measured with a 50$\,$pA AC excitation and zero DC bias, similar to Figure 1d, which explains why the quantum Hall plateaus are eroded by superconducting regions (black spots). Red circles indicate the superconducting pockets shown in Figure 3a-d in the main text. The purple ellipse is the area used in Figure 3d and e to display magnetic interference patterns. Green squares show the locations used to determine the temperature dependence of superconductivity. The yellow rectangle is the area used for Figure 2 in the main text. 	

	Figure S5 shows an alternative representation of the coexistence of superconductivity and the quantum Hall effect for junctions $J_{1-3}$ (panels a, b and c respectively.) Instead of showing two different maps at zero and finite DC bias we plot alternating lines measured at zero DC bias and finite DC bias on the same fan diagram. Even lines are measured at finite bias and show the conventional quantum Hall effect. Odd lines are measured at zero bias and often show a lower differential resistance (darker), which vanishes in superconducting pockets. We cannot extract a clear dependence on junction geometry yet, but signatures of superconductivity are stronger in the shortest junction $J_{1}$ and weaker in the longest, $J_{2}$.
	
	\newpage
	\begin{suppfigure*}[h!]
		\center \label{fig4-1}
		\includegraphics[width=142mm]{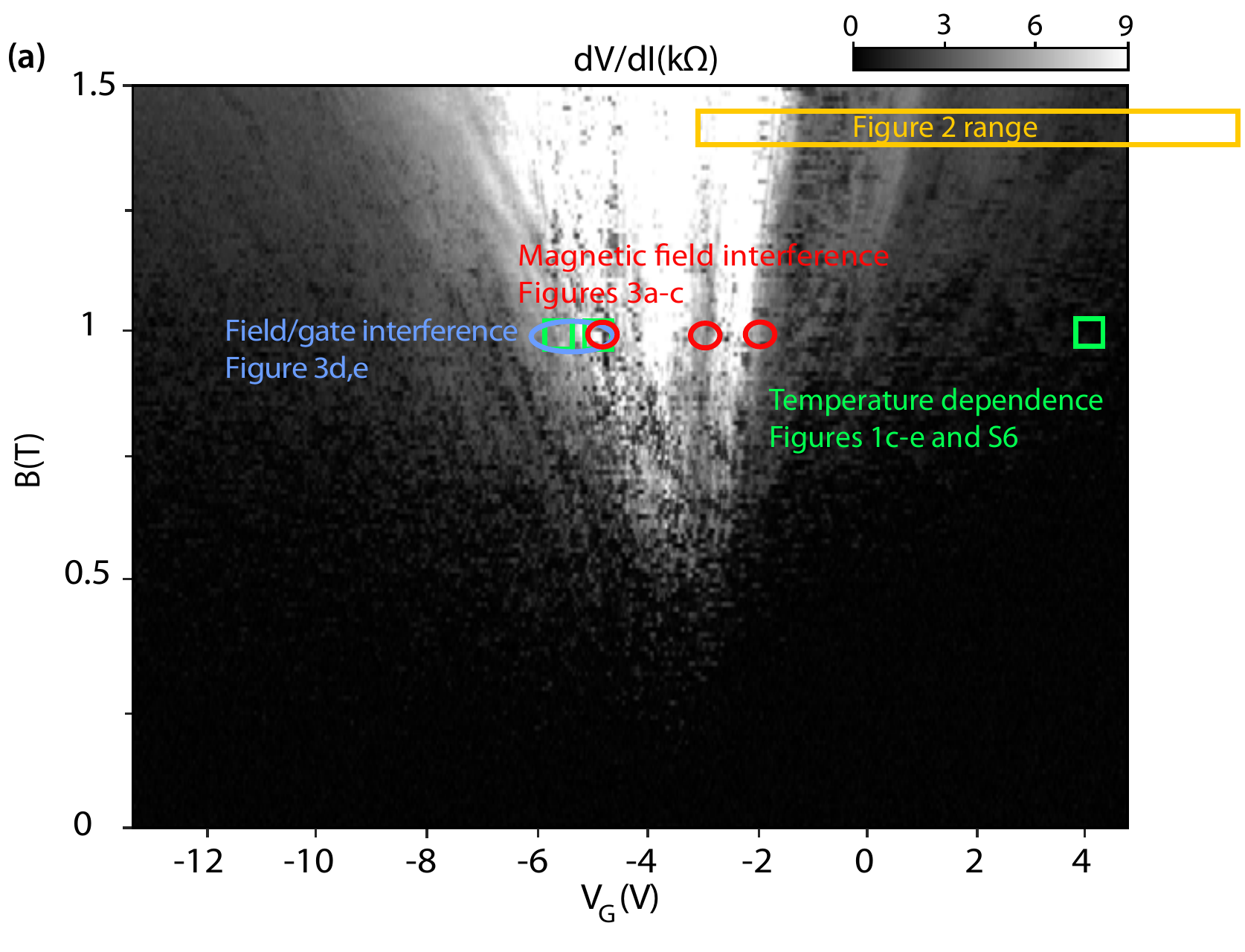}
			\setlength{\abovecaptionskip}{-4pt}
			\caption{Locations of measurements shown on top of the quantum Hall fan diagram of $J_{1}$. The diagram is measured with zero DC bias, to demonstrate the superconducting pockets that erode the QH plateaus (black spots).}
	\end{suppfigure*}
	
	\begin{suppfigure*}[h!]
		\center \label{J1_fan_map}
		\includegraphics[width=142mm]{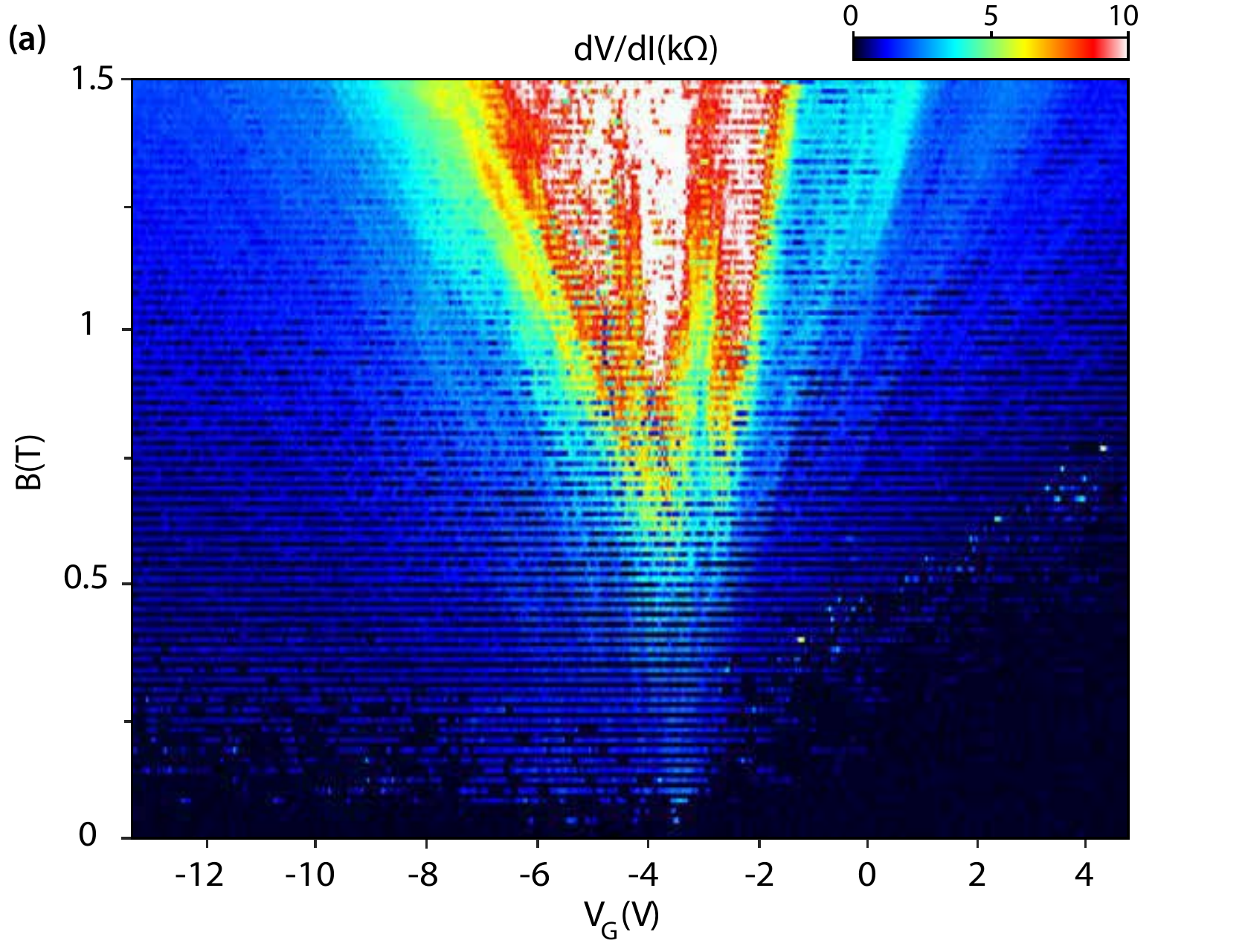}
		\begin{spacing}{1.0}
			    \setlength{\abovecaptionskip}{2pt}
			\caption*{Supplementary Figure 5:(a) Fan diagram for $J_{1}$ showing alternating lines taken at zero and finite DC current.}
		\end{spacing}
	\end{suppfigure*}

	\begin{suppfigure*}[h!]
		\center \label{J1_fan}
		\includegraphics[width=155mm]{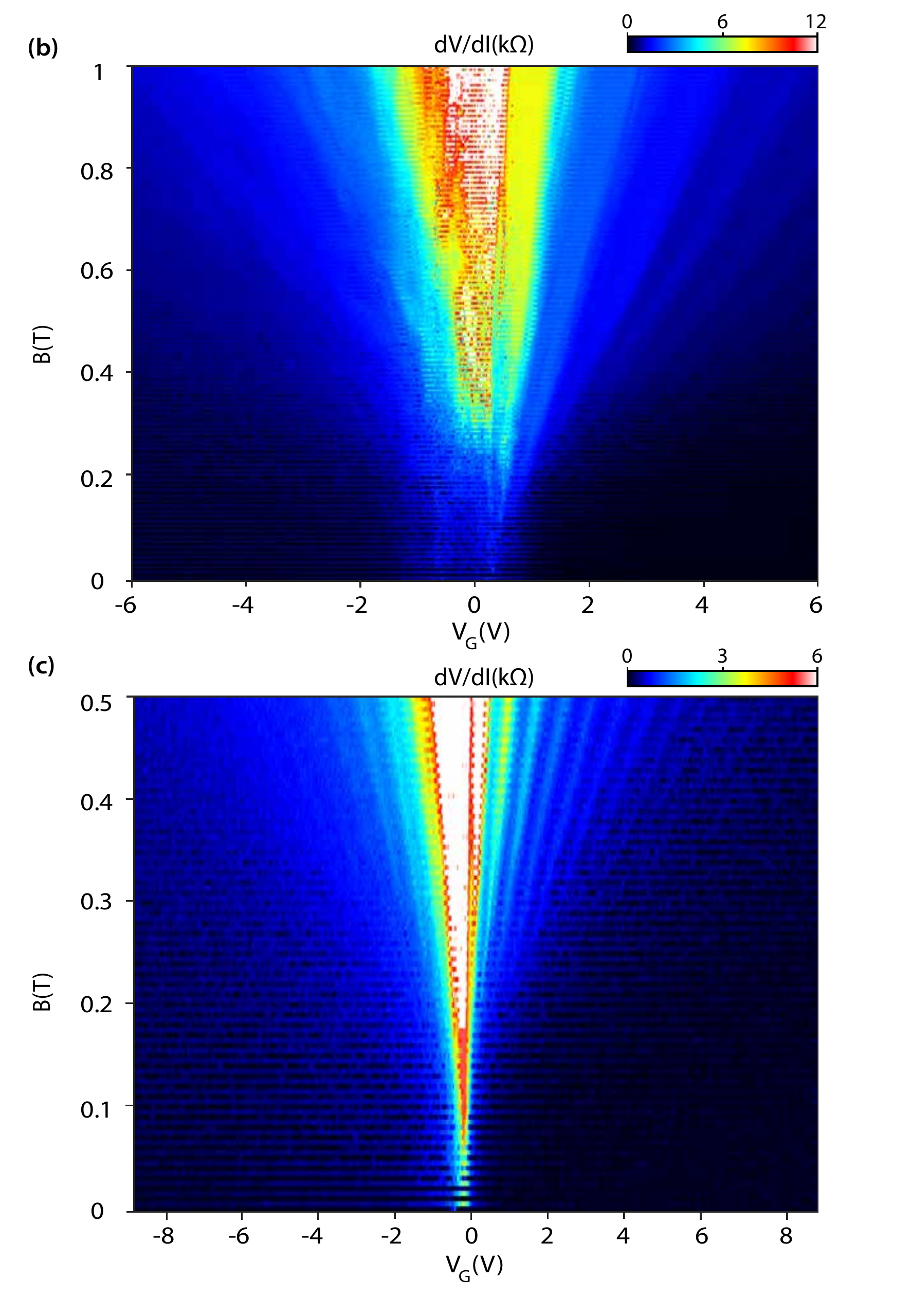}
		\begin{spacing}{1.0}
			\setlength{\abovecaptionskip}{2pt}
			\caption{ (b,c) Fan diagrams for $J_{2-3}$ showing alternating lines taken at zero and finite DC current.}
			\end{spacing}
	\end{suppfigure*}

\clearpage

	\section{Additional temperature dependence of the supercurrent}
	
	We determined the temperature dependence of the differential resistance for several superconducting pockets in the quantum Hall regime. The main purpose of these measurements is to verify that the switching current and the Josephson energy scale directly extracted from the $I$$\,-\,$$V$ curves are meaningful. Indeed, although very small switching currents are sometimes taken to represent the true critical current (which determines the Josephson energy), it is clear that the nA currents, with the Josephson energy in the range of tens of mK, will be strongly affected by thermal fluctuations. In this regime, thermal excitation of the phase causes the phase diffusion; measuring its temperature dependence allows for an independent verification of the Josephson energy.  
	
	\begin{suppfigure*}[h!]
		\center \label{fig5}
		\includegraphics[width=172mm]{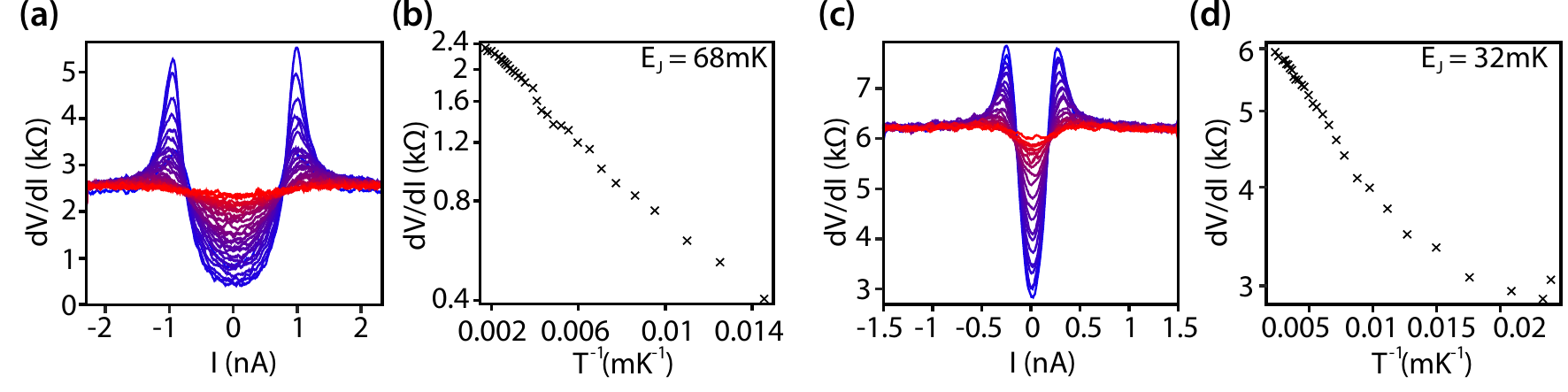}
      
			\caption{(a) Temperature dependence of $dV/dI$, measured on junction J$_{1}$ at $\Vbg$$\,=\,$4$\,$V and $B$$\,=\,$1$\,$T from 60 to 500$\,$mK. (b) Arrhenius plot of the minimum $dV/dI$ vs. $1/T$ on the logarithmic scale. A linear fit yields $E_{J}$$\,\approx\,$70$\,$mK. (c) Temperature dependence of $dV/dI$, measured on junction J$_{1}$ at $\Vbg$$\,=\,$-4.9$\,$V and $B$$\,=\,$1$\,$T from 60 to 500$\,$mK. (d) Arrhenius plot of the minimum $dV/dI$ vs. $1/T$ on the logarithmic scale. A linear fit at hight $T$ yields $E_{J}$$\,\approx\,$30$\,$mK.}	
	\end{suppfigure*}

	Figure S6a shows the temperature dependence of $dV/dI$ vs. $I$, measured on junction J$_{1}$ at $\Vbg$$\,=\,$4$\,$V and $B$$\,=\,$1$\,$T from 60 to 500$\,$mK. An Arrhenius fit to the minimum of $dV/dI$ vs. $T$ yields the Josephson energy scale of approximately 70$\,$mK. This corresponds to a critical current of $I_C \sim 3\,$nA, compared to the measured switching current of $I_S$$\,=\,$1$\,$nA. Indeed, at the temperature of 40 mK, so that $kT/E_{J} \approx 0.5$, the thermal fluctuations are expected to make the measured switching current several times smaller than the true critical current. Additional data measured on a different pocket is shown on Figure S6c,d, taken at $\Vbg$$\,=\,$-4.9$\,$V and $B$$\,=\,$1$\,$T. It yields an extracted Josephson energy of 30$\,$mK ($I_{C} \sim 1\,$nA) for a switching current of $I_S=0.3\,$nA.
	
	\section{Bias and gate dependence of the supercurrent}
	\begin{suppfigure}[h!]
		\center \label{fig7}
		\includegraphics[width=172mm]{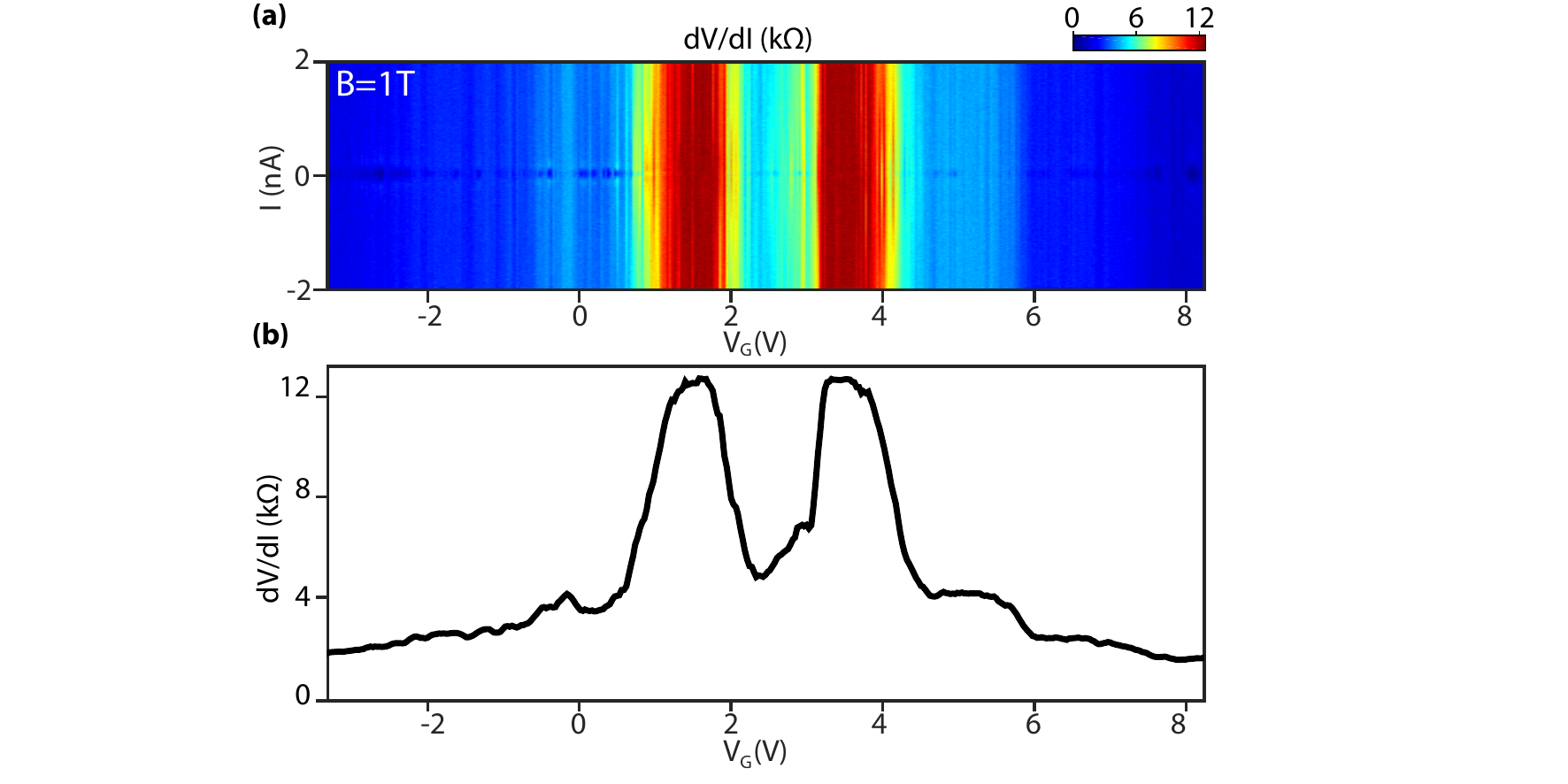}
		\begin{spacing}{1.0}
			\caption{(a) Differential resistance $dV/dI$ measured at 65$\,$mK as a function of bias current $\idc$ and gate voltage $\Vbg$ at $B=1\,$T. Pockets of superconductivity appear close to zero current. (b) Line cut of $dV/dI$ in the same range of $\Vbg$, measured at a finite DC current of 2 nA applied to suppress the supercurrent. Plateaus of quantized resistance are clearly visible at $\nu=4(n+1/2)$ with an integer $n$. }
		\end{spacing}
	\end{suppfigure}
	
	Figure S7a shows $dV/dI (\Vbg, I)$ measured at 1 T and 65 mK for device $J_{4}$. In that regime, $\rc$ is much smaller than the dimensions of the device and the bulk of the graphene sheet is in the quantum Hall regime. Vertical strips of constant resistance in Figure S7a correspond to the quantum Hall plateaus. When applying a DC current of 2$\,$nA, supercurrent is suppressed and plateaus of $dV/dI$($\Vbg$) are observed [Fig. S7b]. Plateaus are better defined in the electron doped regime as a result of the better transmission at the contacts. Regions of suppressed resistance are visible around zero current. In magnetic field, these pockets demonstrate the same types of interference patterns as shown in Figure 3 of the main text. We discuss them in some detail in Figure S8 below.
	
	\section{Additional interference patterns in the quantum Hall regime}
	
	\begin{suppfigure}[h!]
		\center \label{fig8}
		\includegraphics[width=172mm]{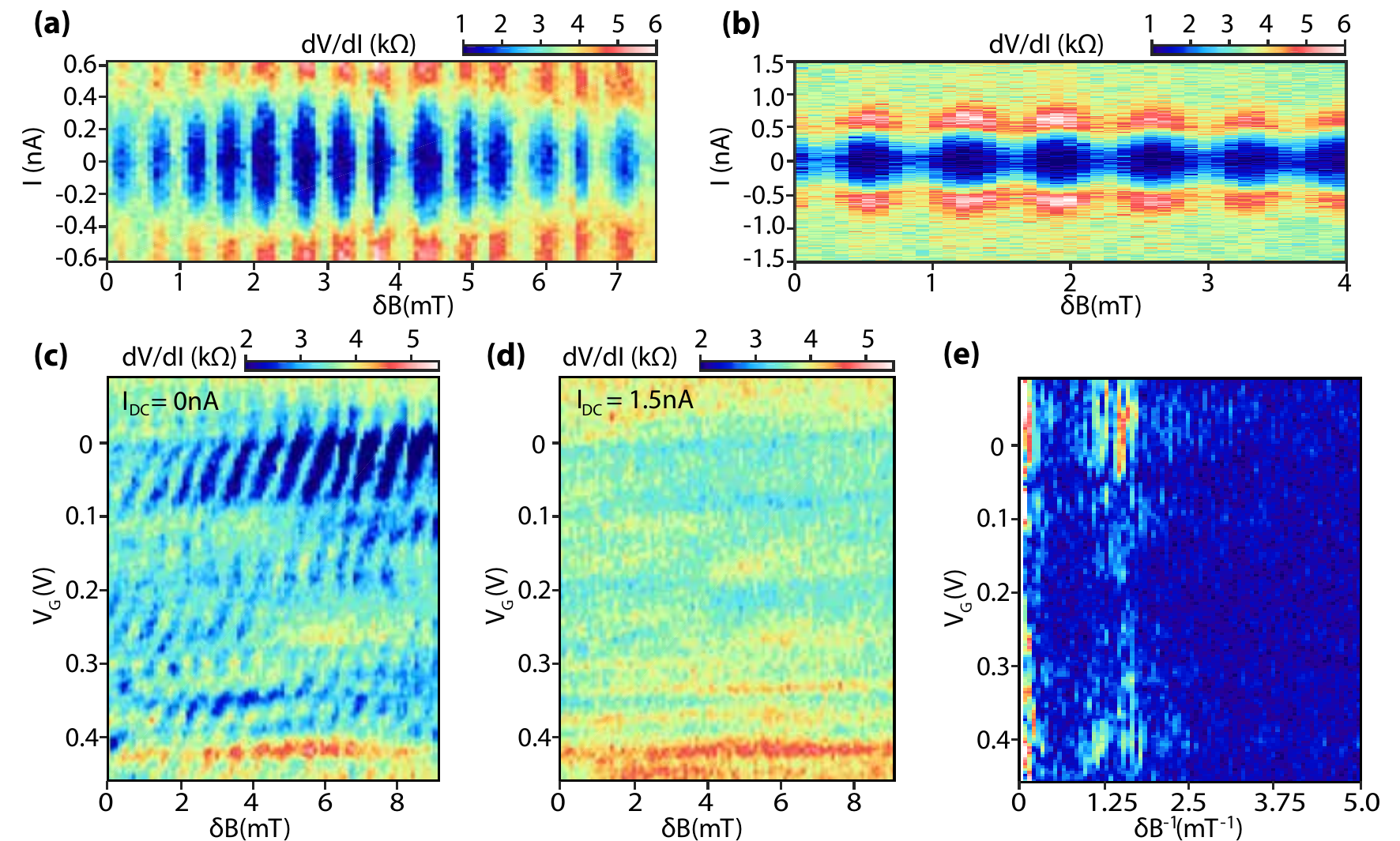}
		\begin{spacing}{1.0}
			\caption{(a,b) Oscillation patterns of $dV/dI(B,I)$ measured on device $J_{4}$ for $\Vbg$$\,=\,$0$\,$V and $\Vbg$$\,=\,$55$\,$mV at magnetic fields close to 1$\,$T. (c,d) $dV/dI(B,\Vbg)$ maps measured on device $J_{4}$ with zero and 1.5$\,$nA DC current close to 1$\,$T. (e) Fourier transform of the map in (c), which shows a band at $1/\delta B$$\,=\,$1.5$\,$mT$^{-1}$, corresponding to the periodicity of $\delta B \approx$ 0.65$\,$mT.}
		\end{spacing}
	\end{suppfigure}
	
	Figure S8 demonstrates the oscillations of the critical current in magnetic field, similar to Figure 3 of the main text, here measured on J$_4$. Figures S8a,b show the supercurrent oscillations in superconducting bubble at $\Vbg$$\,=\,$0$\,$V and $\Vbg$$\,=\,$55$\,$mV around 1$\,$T. 
	Figure S8(c,d) shows another example of the interference pattern as a function of $\Vbg$ and $B$, also measured on J$_4$ with an AC excitation of 50$\,$pA at zero bias (c) and a finite bias of 1.5nA (d). While no field dependence is noticeable in the normal state, an interference pattern with a period of $0.65\,$mT is visible at zero bias. Similar to data shown in the main paper, constant phase contours depend on both $B$ and $V_{G}$. Figure S8e shows the Fourier transform of panel (c). The periodicity of the interference pattern in panel (c) yields the vertical band at the frequency $1/\delta B$$\,=\,$1.5$\,$mT$^{-1}$. This band peaks at $V_{G}$$\,=\,$0 and 80$\,$mV corresponding to major features on panel (c).
	
	\section{Transport in the semiclassical regime}
	
	\begin{suppfigure*}[h!]
		\center \label{fig4-1}
		\includegraphics[width=172mm]{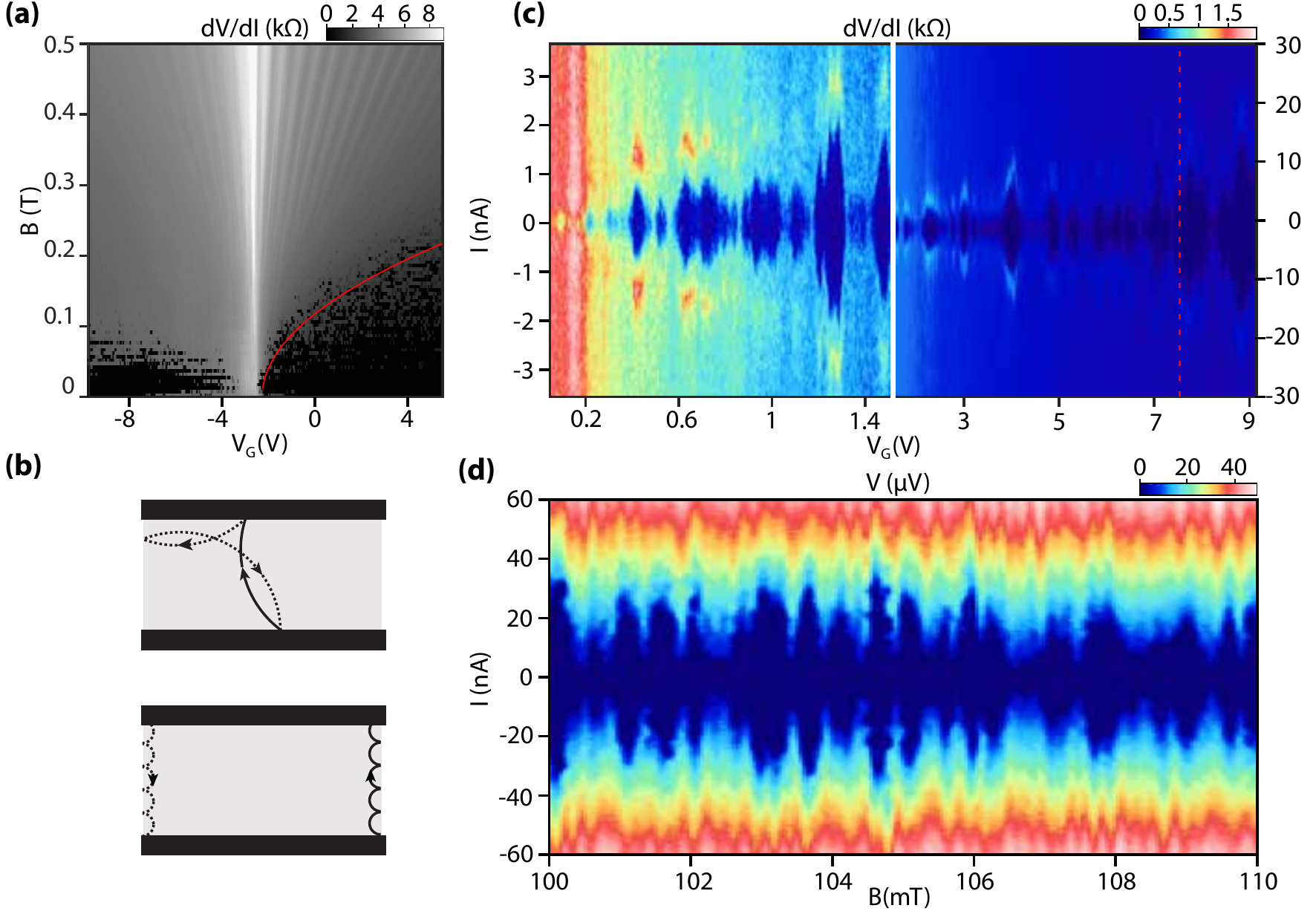}
		\begin{spacing}{1.0}
			\caption{(a) Landau level fan diagram $dV/dI$$\,$($\Vbg,B$) measured at 30$\,$mK with a relatively high AC excitation current of 5$\,$nA, suppressing the supercurrent in the QH regime. (b) Schematics of an Andreev bound state in the semiclassical regime $\rc \gtrsim L$, and in the quantum Hall regime. (c) $dV/dI(\Vbg,I)$ measured on device $J_{3}$ at 200$\,$mT and 35$\,$mK. The left half of the map (from the white line) is a close-up view of $dV/dI$ in a much smaller $\Vbg$ and $I$ range than the right half. The current scales for the left and right sides of the map are respectively indicated on the left and right axes and differ by a factor of 8. The red dashed line correspond to the condition $2 \rc$$\,=\,$$L$. (d) Interference pattern $\vert V(B,I)\vert$ measured at 100$\,$mT in the semiclassical regime at $\Vbg$$\,=\,$8.4V.}
		\end{spacing}
	\end{suppfigure*}
	
	The primary focus of our work was the quantized regime, where the cyclotron orbit $\rc$$\,=\,$$\frac{\hbar \kf}{eB}$ is much smaller than both the mean free path and the device dimensions. Here we discuss magnetotransport in the semiclassical regime $\rc \gtrsim L$, where the deflection of electron hole trajectories is weaker, yet sufficient for the supercurrent not to be uniform in space. 
	
	Figure S9a shows the Landau fan diagram measured on J$_{3}$ with a large excitation current of 5$\,$nA. The excitation current is sufficient to destroy superconductivity in the quantum Hall regime, but a larger supercurrent can persist in the semiclassical regime, under the parabola corresponding to the condition $2\rc$$\,=\,$$L$. This supercurrent has been attributed to Andreev bound states made of closed trajectories connected by multiple elastic and Andreev reflections~$^{[\mathrm{S}4]}$, as schematically shown on Fig. S9b. Figure S9c demonstrates supercurrent in both the semiclassical and the QH regimes. Although the magnitude of the supercurrent changes between the two regimes (notice the change of the vertical scale between the left and the right parts of the map), the transition between the two regimes is gradual, and both parts of the map demonstrate mesoscopic variations of supercurrent as a function of the gate voltage.

	To further illustrate the semiclassical behavior, we demonstrate the voltage across the junction $V(B,I)$ measured in small steps around 100$\,$mT at $\Vbg$$\,=\,$8.4$\,$V (Figure S9d). In this regime, $\rc$$\,\approx\,$920$\,$nm$\,>\,$$L/2=320$nm. Regions of superconductivity are clearly visible, corresponding to $V$$\,=\,$0, with a switching current ranging from 5$\,$nA to 40$\,$nA. These superconducting regions are much less periodic in field and vary greatly in amplitude, in contrast with the QH regime.
	
	We illustrate the dependence of these superconducting pockets on $B$ and $V_G$, by measuring them using the same AC excitation current of 1$\,$nA at a) zero DC current, b) a medium DC current of 6$\,$nA, and c) a large DC current of 100$\,$nA [Fig. S10(a-c)]. Strikingly, at zero bias the sample remains in the superconducting regime throughout the map with very rare spots of finite resistance. At $I_{DC}=6\,$nA, we observe a random patchwork of superconducting regions similar to Ref.$\,$[S4]. This indicates that superconducting regions mostly close and reopen at random fields and densities, but a superconducting current of at least hundreds of pA remains throughout most of this region. At large bias (100$\,$nA), the junction is in the normal state over the entire map, which becomes mostly flat, as expected.

\begin{suppfigure*}[t!]
	\center \label{fig4-2}
	\includegraphics[width=172mm]{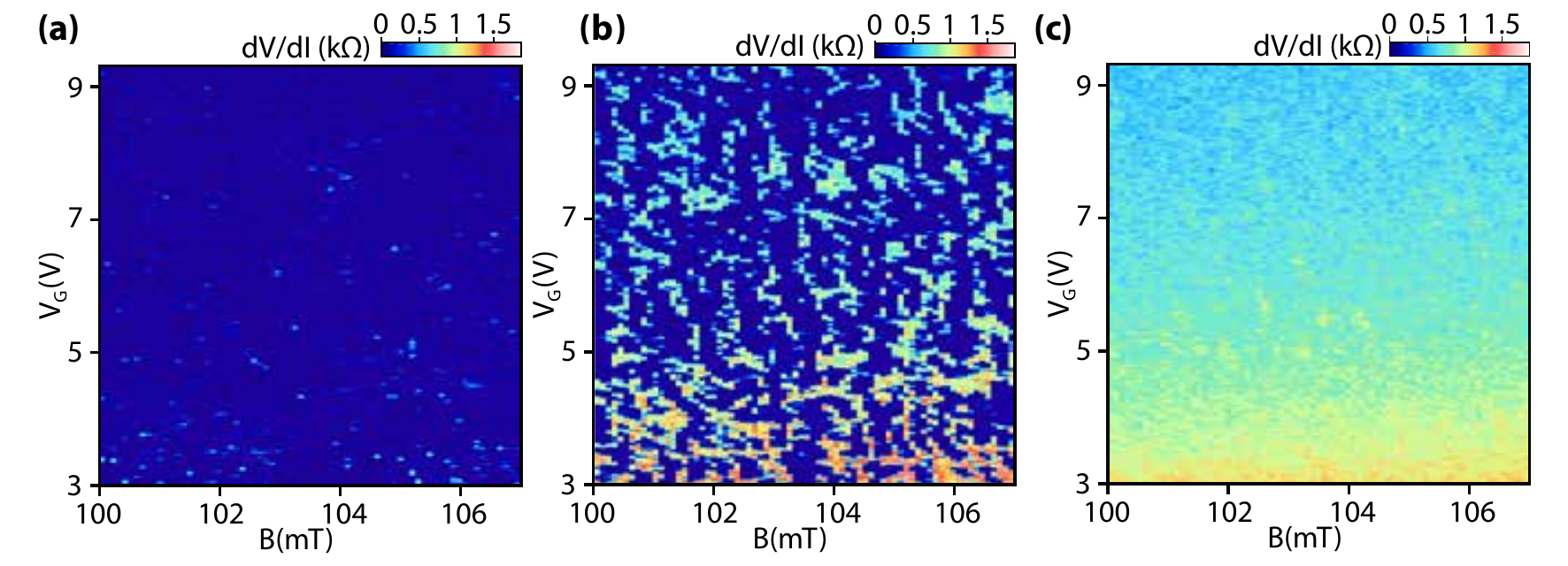}
	\begin{spacing}{1.0}
		\caption{Interference patterns $dV/dI(B, \Vbg)$ measured at a) zero DC current, b) $I_{DC}=6\,$nA, and (c) $I_{DC}=100\,$nA. Supercurrent drops below 6 nA at random values of $B$ and $\Vbg$, but a sub-nA supercurrent persists throughout.}
	\end{spacing}
\end{suppfigure*}

\newpage
	\paragraph{References for the supplementary information}
	
	\begin{itemize}
		\item[S1.] L. Wang, I. Meric, P. Y. Huang, Q. Gao, Y. Gao, H. Tran, T. Taniguchi, K. Watanabe, L. M. Campos, D. A. Muller, J. Guo, P. Kim, J. Hone, K. L. Shepard,  C. R. Dean. One-dimensional electrical contact to a two dimensional material. Science \textbf{342}, 614-617 (2014).
		
		\item[S2.] V. E. Calado, S. Goswami, G. Nanda, M. Diez, A. R. Akhmerov, K. Watanabe, T. Taniguchi, T. M. Klapwijk, L. M. K. Vandersypen. Ballistic Josephson junctions in edge-contacted graphene, Nature Nano. \textbf{10}, 761-764 (2015).
		
		\item[S3.] N. Mizuno, B. Nielsen, X. Du. Ballistic-like supercurrent in suspended graphene Josephson weak links. Nature Commun. \textbf{4}, 2716 (2013).
		
		\item[S4.] M. Ben Shalom, M. J. Zhu, V. I. Fal'ko, A. Mishchenko, V. Kretinin, K. S. Novoselov, C. R. Woods, K. Watanabe, T. Taniguchi, A. K. Geim.  $\&$ Prance, J. R. Proximity superconductivity in ballistic graphene: from Fabry-Perot oscillations to random Andreev states in magnetic fields. Preprint at: http://arxiv.org/abs/1504.03286.
		
		\item[S5.] M. T. Allen, O. Shtanko, I. C. Fulga, J. I. J. Wang, D. Nurgaliev, K. Watanabe, T. Taniguchi, A. R. Akhmerov, P. Jarillo-Herrero, L. S. Levitov, A. Yacoby. Visualization of phase-coherent electron interference in a ballistic graphene Josephson junction. Preprint at: 
		http://arxiv.org/abs/1506.06734 (2015).
		
		\item[S6.] M. Tinkham, Introduction to Superconductivity, New York: McGraw-Hill (1996).
		
		\item[S7.]P. A. Rosenthal, M. R. Beasley, K. Char, M. S. Colclough, and G. Zaharchuk. Flux focusing effects in planar thin-film grain-boundary Josephson junctions, Applied Physics Letters 59, 3482 (1991).
		
	\end{itemize}

\end{document}